 \documentstyle[preprint,prl,aps]{revtex}
\begin{document}
\draft

\title{
Impurity Energy Level Within The Haldane Gap
}
\author {Wei Wang,${}^1$ Shaojin  Qin,${}^2$
Zhong-Yi Lu,${}^{3}$ Lu  Yu,${}^{2,4}$ and
Zhaobin  Su ${}^2$}
\address{
${}^1$ Physics Department of Peking University,
 Beijing 100871, China \\
${}^2$Institute of Theoretical Physics,
 Academia Sinica,
  Beijing 100080, China\\
${}^3$International School for Advanced Studies, 34013 Trieste, Italy\\
${}^4$International Centre for Theoretical Physics,
PO. Box 586, 34100 Trieste, Italy}
%\date{June,1995}
\maketitle
\begin{abstract}
An impurity bond $J{'}$ in a periodic 1D antiferromagnetic,
spin 1  chain with exchange $J$ is considered. Using
the numerical density matrix renormalization group method, we find
an impurity energy level in the Haldane gap,
corresponding to a bound state near the impurity bond. When
$J{'}<J$ the level changes gradually from the edge of the
Haldane gap to the ground state energy as the deviation
$dev=(J-J{'})/J$ changes from 0 to 1.  It seems that
there is no threshold. Yet, there is a threshold
when $J{'}>J$. The  impurity level appears only
when the deviation $dev=(J{'}-J)/J{'}$ is greater
than $B_{c}$, which is near 0.3 in our calculation.
\end{abstract}

\maketitle

\section{introduction}

    The Heisenberg model of $1D$ antiferromagnetic(AF)
chains is :
\begin{equation}
    $$H=J\sum\limits_{i}^{}{\bf S}_{i}\cdot{\bf S}_{i+1}$$
\end {equation}
with $J>0$, and  ${\bf S}_{i}$ is a  quantum spin. Up till now,
the exact solution of this Hamiltonian is
available only when $s=1/2$. The solution is obtained by the
Bethe {\sl ansatz}\cite{bethe}, which shows there is
no gap in the energy spectrum. Then, what is the
situation when $s=1$ or other integer? Some years ago,
Haldane made his famous conjecture\cite{haldane2} that
$1D$ AF chains of integer spin have a gap, while those of
half integer spin have a gapless spectrum. Since then, a lot of work has
been done. In particular, a Valence-Bond-Solid (VBS) picture was proposed
by Affleck, Kennedy, Lieb, and Tasaki(AKLT) \cite{aklt3,aklt4}
to interprete the
ground state of the integer spin AF chains.
The proposed picture agrees quite well with both
 experimental\cite{exp5} and  numerical\cite{num6}
studies.

    An important issue of integer spin AF chains is the
doping effect which involves
some fundamental many-body quantum problems\cite{lu7}.
DiTusa {\sl et al}\cite{ditusa} have done doping experiments
with ${ Zn}$ or ${Ca}$ in
$Y_{2}BaNiO_{5}$, which contains 1D AF chains of
$Ni-(O)-Ni-(O)-\cdots$. In the ${Zn}$ doping case,
the non-magnetic ${Zn}^{2+}$ ion substitutes ${Ni}^{2+}$  to
sever the  AF chain giving rise  to finite length effect.
On the other hand, in the ${Ca}$ doping  case
(${Ca}^{2+}$ substituting the off-chain
atom $Y^{3+}$), holes are introduced in the oxygen orbitals
along the $Ni-O$ chain, modifying the superexchange
interaction and producing an impurity state inside the Haldane
gap, as seen by neutron scattering experiments.
The latter case can be represented by changing $J$ to $J{'}$
at the impurity bond.
We focus  on this case in this Brief Report. Thus we
write down our Hamiltonian as:
\begin{equation}
  $$H=J\sum\limits_{i}^{}{\bf S}_{i}\cdot{\bf S}_{i+1}
       +(J{'}-J){\bf S}_{0}\cdot{\bf S}_{1}, $$
\end{equation}
 where the impurity bond is put between sites 0 and 1.

   Recently a calculation of the dynamic structure factor based
on the Schwinger-boson approach was performed for the bond-doping
case\cite{lu7}, which first indicated the existence of
a threshold for the  bond-coupling deviation.  According to this calculation,
a bound state is induced by the impurity only if the deviation
exceeds this threshold, and the impurity level is located in the
middle of the Haldane gap, almost independent of bond-coupling
when temperature is much lower than the Haldane gap.

   The recently developed numerical method of density matrix
 renormalization group (DMRG)\cite{white1,white9}
has achieved a great success in calculating the low energy
spectrum of 1D Heisenberg AF chains\cite{white10}, so it's
interesting and natural to use this method to study the
doping effect on the Haldane gap systems. Such a calculation was
carried out
by Sorenson and Affleck (SA)\cite{sorenson8}. They reported
the existence of the threshold, but the impurity energy level
they obtained changes
gradually as the strength of impurity bond changes.  The existence
of the threshold is in agreement with
the Schwinger boson calculation\cite{lu7}, while the gradual change of the
impurity level is in disagreemnet with the latter.
However, these authors have used the open boundary conditions
and the way they added new sites may affect the
bound state itself (See the discussion in the next Section).
To clarify this issue we have decided
to devise a more elaborate scheme to carry out the DMRG calculations.

\section {DMRG method}

      The DMRG method has been developed  by S.R. White {\sl et al}
\cite {white1,white9} in recent years. Generally, the standard pattern
looks like this:

$$
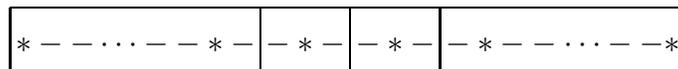
\begin{tabular}{|c|c|c|c|} \hline
 $ \ast -- \cdots -- \ast - $ & $ - \ast - $
 & $ - \ast - $ & $ - \ast -- \cdots -- \ast $  \\  \hline
\end{tabular}
$$

%                  _________       _______
%                  |*--*--*-|-*--*-|-*--*-* |
%                   --------         -------

    During each iteration, the chain (superblock) is divided
into $4$ blocks, where the left-most and the right-most (which can
be named as old blocks) are the same for the symmetry consideration.
The two middle  blocks contain only two sites. After each run,
the left and right halves  of the superblock are optimized
to form  new blocks which are used for the next iteration. This
is, however, not good in the presence of an impurity bond. At first
glance it seems right to put
 the impurity bond  in the middle of the
chain (this is what SA have done
\cite{sorenson8}), but this configuration may introduce some
artificial   effects. For
example, the two middle sites which were optimized with the
impurity bond between them, are used in later calculations
without  impurity being close to them. When the chain is very long,
the impurity bond effect
should go to zero during iterations (since there is only one impurity bond),
 but using this pattern, the
effect cannot become small as iterations go on.

     To calculate the impurity level of the periodic chain,
we do our numerical work using the following pattern:

$$
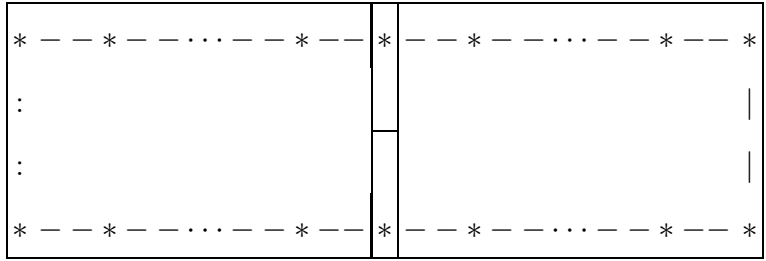
\begin{tabular}{|cc|c|cc|}  \hline
 $ \ast $ & $-- \ast -- \cdots -- \ast --$ & $\ast$ &
 $-- \ast -- \cdots -- \ast --$ & $\ast$  \\
   : &          &   &         & $|$  \\
 \cline{3-3}
   :  &         &   &         & $|$  \\
 $ \ast $ & $-- \ast -- \cdots -- \ast --$ & $\ast$ &
 $-- \ast -- \cdots -- \ast --$ & $\ast$  \\  \hline
\end{tabular}  \\
$$
%                   *--*--*--*--*--*--*--*
%                   :                    |
%                   :                    |
%                   *--*--*--*--*--*--*--*

   The impurity bond is placed at the left-hand side, and we add
two sites each time with  one each at or near the center of
the upper and lower parts of the above periodic chain. Using this
pattern, one can avoid the difficulties mentioned above. We preserve
$m=100$ states in each run and discard the rest. Since $S_{z}$
is a good quantum number, we use it to reduce our Hamiltonian
matrix.

    Shank transformation has been demonstrated as a powerful
extrapolation method to get Haldane gap precisely\cite{golinelli}.
We use it to extrapolate  our results to the infinite length limit.
To verify the credibility of this method, we also use Aitkens'
transformation\cite{aitken} to extrapolate our finite chain data.
They both give very similar values (usually the difference
is less than $10^{-3}$). Compared with the previous
conclusion in special cases, our results show reasonable
precision. So we believe that our value is correct at least up
to $10^{-2}$.

\section{ numerical result}

    We use VBS picture\cite{aklt3,aklt4} to give some
intuitive explanation of our calculation results, although this
picture is not quite precise. According to this picture,
when considering
the low excitation energy spectrum, the spin $1$ of each
site can be regarded as two spin $1/2{'}s$ combined
together to form a triplet $S=1$ state, yet the two
$1/2$  spins of the nearest sites form a Valence-Bond,
or, a singlet state.  Therefore, the ground state of
an open chain can be labelled by the sum of the two edge
sites $1/2$ spins.  The  ground state of the 1D AF chain
is  four-fold degenerate,  with one state of
${\bf S}_{total}=0$,
where the two $1/2$ spins of each edge form a singlet,
 and three states of  ${\bf S}_{total} =1$ ,
where the two $1/2$ spins form a triplet.
The first excited states are the quintlet states with
${\bf S}_{total}=2$, which lie above the ground state by
the famous Haldane gap. Above them, the continuous
spectrum starts. For a periodic chain, VBS picture also
holds, but the triplet state rises up to the edge of
the continuous energy spectrum, leaving the ground state
nondegenerate.

      The impurity state of the periodic chain, is just
the ${\bf S}_{total}=1$ triplet state. When $J{'}/J=0$,
the periodic chain becomes  an open chain, and this triplet
will merge with the  ${\bf S}_{total}=0$ ground state
in the infinite chain length limit. Thus the impurity energy
level is  at $0$. On the contrary, when $J{'}/J=1$, the
periodic chain recovers, so  the "impurity" level will
rise up to merge with the Haldane gap.

      Because of the $SO(3)$ symmetry of the Heisenberg
Hamiltonian , the ${\bf S}_{total}=1$ triplets are degenerate,
so we choose the state  $S_{z}=1$ as a representative of
them. Also we set the zero of the energy spectrum as the
ground state energy.

     First we consider  the case $0<J{'}<J$, which means the super-
exchange is smaller due to doping, yet  it does not change
the AF nature.  The calculated result is presented in Figure
1, from which  we see that the impurity level increases linearly
near $dev=1$($dev=(J-J{'})/J$), which is in accordance with
SA's perturbation argument\cite{sorenson8}. However,
compared with SA's work\cite{sorenson8}, our numerical
result favors the conclusion that there is no threshold.
As seen from the figure, the impurity energy appears as soon as
$dev\not=0$, and shows a quadratic behavior near $dev=0$.
Kaburagi {\sl et al} \cite{kaburagi} has investigated the impurity
bond effect in terms of domain wall exitations. Their variational
results show the linear behavior near $dev=0$ and
quadratic behavior near $dev=1$.  They also perform
diagonalizations by Lanczos method for $N=12$ and $13$
chains to support the variational results.
%These features  agree with the variational calculation
%of Kaburagi's\cite{kaburagi}

      When $J{'}>J$, the result is shown in Figure 2, where
the impurity level vs $dev=(J{'}-J)/J{'}$ is plotted. This
time, a threshold $B_{c}$ appears near 0.3, i.e., only when
$dev>B_{c}$ an impurity localized state will appear. Again
the level rises linearly near $dev=1$, in
agreement with the results of SA\cite{sorenson8}, but with
a different $B_{c}$ (they gave  $B_{c}=0.5$).  The case
when $J{'}/J \rightarrow \infty$ ($dev\rightarrow 1$) can
be understood easily as follows:  The two spins of the impurity
bond form a singlet state because of the strong coupling,
breaking the periodic chain of length L to make an open chain
of length L-2. So, as following from our calculation, the difference
between the impurity level and the ground state will go to zero
in  this limit.
According to the finit chain results, the impurity level
when $0<dev<B_{c}$ is greater than the level when $dev=0$, and the
difference between them become smaller when the chain length
increases,so we believe there is a threshold when the length
goes to infinity, instead of the exponential or other behavior
of the energy near $dev=0$ .

     Moreover, to ensure that the impurity level means a local state
near
the impurity bond, we plot $<S_{zi}>$ of the periodic chain
in Figure 3 ($J{'}<J$). When $J{'}/J=0$($dev=1$), the "boundary
$1/2$ spin" of open chain appears, just as the VBS picture assumes
\cite{aklt3}. When $dev$ goes from  $1$ to $0$, the state
becomes more and more delocalized. Figure 4 shows the case of
$J{'}>J$, where the two sites at the middle form a singlet state,
while the sites  near them show the "boundary $1/2$ spin", too, when
$dev \rightarrow 1$.

     Figure 5 and 6 give the Shank transformation results of
$|{dE}/{dDev}|$. There is obviously no abrupt change.

\section{conclusion}

     Using an elaborate numerical DMRG method, we have investigated
the bond doping effect of 1D AF chain. A threshold $B_{c}$
exists only when $J{'}>J$. When $J{'}<J$ or $J{'}>J$ but
$dev< B_{c}$, no impurity state is induced. When $dev>B_{c}$
and $J{'}>J$, the impurity  level  within the
Haldane gap corresponds to a local state near the impurity bond,
and the
level changes gradually when $dev$ runs from $B_{c}$ to 1.
The discrepancy of the present study with the Schwinger boson
calculation\cite{lu7} is not fully understood at present. One possiblity
is that the Schwinger boson representation introduces an additional
symmetry (which sets the impurity level at the middle of the Haldane gap),
not inherent to the original quantum spin systems.

     Our calculation is supported by LSEC, CCAST and
Laboratory of Computational Physics,IAPCM.
 We appreciate their assistance. After completion of
 our calculations we received a preprint by Xiaoqun Wang and S. Mallwitz,
which reports essentially the same results. We thank Dr. Wang for sending
us the preprint.

\begin{figure}
%\centerline{\psfig{figure=Fig/ggl.ps,height=3.5inch,clip=}}
\caption{
Impurity energy level vs impurity $dev=(J-J{'})/J$,
when $0<J{'}<J$. The chain is periodic, and the dashed
line indicates the Haldane gap.
}
\end{figure}

\begin{figure}
%\centerline{\psfig{figure=Fig/ggg.ps,height=3.5inch,clip=}}
\caption{
Impurity energy level vs impurity $dev=(J{'}-J)/J{'}$,
when $J{'}>J$. The chain is periodic, and the dashed line
indicates the Haldane gap.
}
\end{figure}

\begin{figure}
%centerline{\psfig{figure=Fig/szlt1.ps,height=3.5in,clip=}}
\caption{
$<S_{zi}>$  of the impurity state as function of chain
index $i$ when $J{'}<J$. The star is for the
case of $dev=1$, while the triangle  is for $dev=0.6$, and
the square for  $dev=0$ ($dev=(J-J{'})/J$). The impurity
bond is placed  between site 15 and 16.
}
\end{figure}

\begin{figure}
%centerline{\psfig{figure=Fig/szgt1.ps,height=3.5in,clip=}}
\caption{
$<S_{zi}>$ of the impurity state as funtion of chain
index $i$ when $J{'}>J$.  The  star is for the
case of $dev=0.999$, while the triangle is for  $dev=0.7$,
and the square for $dev=0$ ($dev=(J{'}-J)/J{'}$). The impurity
bond is placed between site 15 and 16.
}
\end{figure}

%\begin{figure}
%%centerline{\psfig{figure=Fig/dglt1.ps,height=3.5in,clip=}}
%\caption{
%$|{dE}/{dDev}|$ vs $Dev$ when $J{'}<J$. No abrupt
%change appears.
%}
%\end{figure}

%\begin{figure}
%%centerline{\psfig{figure=Fig/dggt1.ps,height=3.5in,clip=}}
%\caption{
%$|{dE}/{dDev}|$ vs Dev, when $J{'}>J$. Also, no abrupt
%change appears.
%}
%\end{figure}


\begin{references}

\bibitem{bethe}
H.A. Bethe,   Z. Phys. {\bf 71}, 205 (1931).
\bibitem{haldane2}
F.D. Haldane,  Phys. Lett. ${\bf 93A}$,  464 (1983);
\prl {\bf 50},1153 (1983).
\bibitem{aklt3}
I. Affleck,T. Kennedy, E.H. Lieb and H.Tasaki, \prl
{\bf 59}, 799 (1987).
\bibitem{aklt4}
I. Affleck, T. Kennedy, E.H. Lieb and H. Tasaki,
Commun. Math. Phys.{\bf 147}, 431 (1992).
\bibitem{exp5}
See,e.g.,  J.P.Renard {\sl et al},  Europhys.
Lett. {\bf 3},945 (1987);
W. Lu,{\sl et al} \prl {\bf 67}, 3538 (1988);
L.C. Brunel {\sl et al}
\prl, {\bf 69},1699 (1992).
 \bibitem{num6}
See, e.g.  O. Golinelli {\sl et al},
\prb {\sl 50},3037(1994);
J.B. Parkinson and J.C. Bonner,
\prb {\bf 32},4703(1985);
S. Liang,   \prl  {\bf 64},1597 (1990);
M.P. Nightingale and H.W. J.Blote,
\prb {\bf 33}, 659 (1986).
\bibitem{lu7}
Zhong-Yi Lu, Zhao-Bin Su,and Lu Yu,
\prl {\bf 74}, 4297 (1995).
\bibitem{ditusa}
J.F. DiTusa {\sl et al}, \prl {\bf 73},1857 (1994).
\bibitem{white1}
S.R. White,  \prb {\bf 48},10345 (1993).
\bibitem{white9}
S.R. White and R.M. Noack,   \prl {\bf 68}, 3487 (1992).
\bibitem{white10}
S.R.White,   \prl {\bf 69}, 2863 (1992);
S.R.White and D.A.Huse,  \prb {\bf 48}, 3844 (1993).
\bibitem{sorenson8}
E. S. Sorenson and I. Affleck,  preprint 1994.
\bibitem{golinelli}
O.Golinelli {\sl et al}, \prb {\bf 50}, 3037 (1994).
\bibitem{aitken}
see,e.g.,C.Bresinski and M.R.Zaglia {\sl Extrapolation
Methods}(Elsevier,New York,1991).
\bibitem{kaburagi}
M.Kaburagi {\sl et al}, J. of Phys. Soc. of Japan
{\bf 62},1848 (1993).
\end{references}
\end{document}